\documentstyle[epsf,amsmath,amssymb,graphicx,pramana]{ias}
\newcommand{\abbrev}{\small}

\newcommand{\eqn}[1]{Eq.\,(\ref{#1})}
\newcommand{\fig}[1]{Fig.\,\ref{#1}}
\newcommand{\tab}[1]{Table\,\ref{#1}}

\newcommand{\dd}{{\rm d}}

\newcommand{\lhc}{{\abbrev LHC}}
\newcommand{\sm}{{\abbrev SM}}
\newcommand{\mssm}{{\abbrev MSSM}}
\newcommand{\susy}{{\abbrev SUSY}}

\newcommand{\lo}{{\abbrev LO}}
\newcommand{\nlo}{{\abbrev NLO}}
\newcommand{\nnlo}{{\abbrev NNLO}}
\newcommand{\nklo}[1]{{\abbrev N$^{#1}$LO}}
\newcommand{\msbar}{\mbox{$\overline{\mbox{\abbrev MS}}$}}
\newcommand{\muF}{\mu_{\rm F}}
\newcommand{\muR}{\mu_{\rm R}}

\newcommand{\ffs}[1]{{\abbrev #1-FS}}
\newcommand{\wbf}{{\abbrev WBF}}
\newcommand{\qcd}{{\abbrev QCD}}
\newcommand{\GFermi}{G_{\abbrev F}}

\title{Standard and SUSY Higgs production at the LHC}
\author{Robert Harlander}
\address{Fachbereich C -- Theoretische Physik, Bergische Universit\"at
  Wuppertal, 42097 Wuppertal, Germany}
\abstract{
Recent theoretical developments concerning Higgs production at the Large
Hadron Collider are reviewed, both in the Standard Model and in the
\mssm{}. Emphasis is put on the inclusive and exclusive cross sections
for gluon fusion, as well as on the associated production with bottom
quarks.}

\begin{document}
\maketitle

\section{Introduction}

The upcoming searches for and anticipated studies of Higgs bosons at the
Large Hadron Collider (\lhc{}) have been the driving force for
interesting theoretical developments for many years. Quite a number of
observables and input quantities to experimental analyses are meanwhile
known to high precision. In particular, the most important Higgs
production cross sections are under good theoretical control: the
next-to-leading order (\nlo{}) predictions for associated $t\bar tH$
production~\cite{Beenakker:2001rj:2002nc,Dawson:2002tg,Dawson:2003zu}
and weak boson fusion \cite{Figy:2003nv}, as well as the
next-to-next-to-leading order (\nnlo{}) cross sections for gluon
fusion~\cite{Harlander:2002wh,Anastasiou:2002yz,Ravindran:2003um} and
Higgs Strahlung~\cite{Brein:2003wg,Ciccolini:2003jy} all exhibit nicely
converging perturbative series and only a moderate dependence on the
renormalization and factorization scale.

This brief overview should be considered as a report on recent
developments concerning higher order calculations for neutral Higgs
production at the \lhc{}.  For more comprehensive surveys, we refer the
reader to some recent reviews (see, e.g.,
Refs.\,\cite{Djouadi:2005gi:2005gj,Buscher:2005re}).

\section{Gluon fusion}
Higher order corrections to the gluon fusion process have been of great
interest for many years now. The main reason is, of course, that it is
one of the most important discovery channels for Higgs bosons at the
\lhc{}, and that therefore its cross section has to be under good
theoretical control. In fact, the \nlo{} radiative corrections turned
out to be very
large~\cite{Dawson:1990zj,Djouadi:1991tk,Graudenz:1992pv}, amounting to
an increase of up to 100\% with respect to the leading-order (\lo{})
value. In addition, the \nlo{} corrections did not lead to a decrease of
the renormalization and factorization scale dependence (when measured in
absolute rather than relative values of the cross section). Due to these
issues in the theoretical prediction, the $K$-factor was often neglected
in experimental analyses.

A lesson learned from the \nlo{} corrections was that the gluon-Higgs
interaction seems to be approximated very
well~\cite{Kramer:1996iq,Spira:1997dg} by an effective Lagrangian
\begin{equation}
\begin{split}
{\cal L}_{ggH} &= -\frac{H}{4v}\,C(\alpha_s)\,G_{\mu\nu}^aG_a^{\mu\nu}\,,
\label{eq::leff}
\end{split}
\end{equation}
if the \lo{} top mass dependence of the cross section is factored out.
In \eqn{eq::leff}, $v=246$\,GeV and $C(\alpha_s)$ is the Wilson
coefficient which is meanwhile known through
$\alpha_s^5$~\cite{Schroder:2005hy,Chetyrkin:2005ia}. This observation
allowed to tackle the \nnlo{} calculation on the basis of
\eqn{eq::leff}~\cite{Harlander:2000mg,Harlander:2001is,Catani:2001ic}.
The full \nnlo{}
corrections~\cite{Harlander:2002wh,Anastasiou:2002yz,Ravindran:2003um}
exhibited the features of a well-behaved perturbative series: they are
significantly smaller than the \nlo{} corrections, and also the scale
dependence reduces to an acceptable level.

Progress concerning higher order corrections to the gluon fusion process
has been made in various respects. On the one hand, the validity of the
perturbative prediction for the total cross section has been confirmed
by the evaluation of effects that go beyond \nnlo{}. On the other hand,
various resummations for kinematical distributions of the Higgs boson
have been carried out. And finally, a fully differential partonic Monte
Carlo program, valid through \nnlo{}, has been developed. Let us discuss
these topics in more detail in what follows.

{\it Inclusive Higgs production.}\quad With the higher order \qcd{}
effects for gluon fusion being quite sizable, one may wonder about the
reliability of the fixed order \nnlo{} prediction. To answer this
question, one may try to identify and resum the dominant terms to the
inclusive cross section. In fact, it had been realized long ago that
soft gluon radiation contributes significantly to the total
rate~\cite{Kramer:1996iq,Harlander:2001is,Catani:2001ic}.  However, how
well the cross section is approximated by this contribution alone
depends strongly on the way the ``soft limit'' is defined. To
see what we mean by this, consider the general expression for the
hadronic cross section $\sigma(s)$ in terms of parton densities $\phi_i$
($i=q,\bar q,g$) and the partonic cross section $\hat\sigma(\hat s)$:
\begin{equation}
\begin{split}
\sigma(s) &= \int_0^1\dd x_1\int_0^1\dd x_2
\phi_i(x_1)\phi_j(x_2)\hat\sigma_{ij}(x_1x_2 s)\,.
\end{split}
\end{equation}
One way to define the soft limit is to expand $\hat \sigma$ in
the limit $x\equiv M_H^2/\hat s \to 1$:
\begin{equation}
\begin{split}
\hat\sigma_{ij}(\hat s) &=\sigma_0\left( a\,\delta(1-x) + \sum_{k\geq
  0}b_k{\cal D}_k(x) + \cdots\right)
\label{eq::soft}
\end{split}
\end{equation}
where
\begin{equation}
\begin{split}
{\cal D}_k(x) \equiv \left[\frac{\ln^k(1-x)}{1-x}\right]_+
\end{split}
\end{equation}
and the dots denote formally subleading terms which are
dropped. However, one may equally well trade a factor of $x$ between the
partonic cross section and the parton density functions, and rather
expand $\hat \sigma(\hat s)/x$ around $x=1$. This will lead to significantly
different numerical results for the hadronic cross section
$\sigma$~\cite{Harlander:2001is,Catani:2001ic}.

A definition of the soft limit that seems to approximate the full
\nnlo{} cross section very well has been found in
Ref.\,\cite{Catani:2003zt}: one transforms $\hat\sigma(\hat s)/x$ from
$x$ space to Mellin moment space (``$N$ space''), and drops all terms
that are of order $1/N$.  The terms $\alpha_s^n\ln^kN$ were then
resummed to next-to-next-to-leading logarithmic accuracy ({\abbrev
NNLL}), with a rather moderate numerical impact~\cite{Catani:2003zt}. This
indeed indicates the aforementioned stability of the \nnlo{} expression.

\begin{figure}
  \begin{center}
    \leavevmode
    \begin{tabular}{c}
      \includegraphics[bb=0 0 650 420,width=30em]{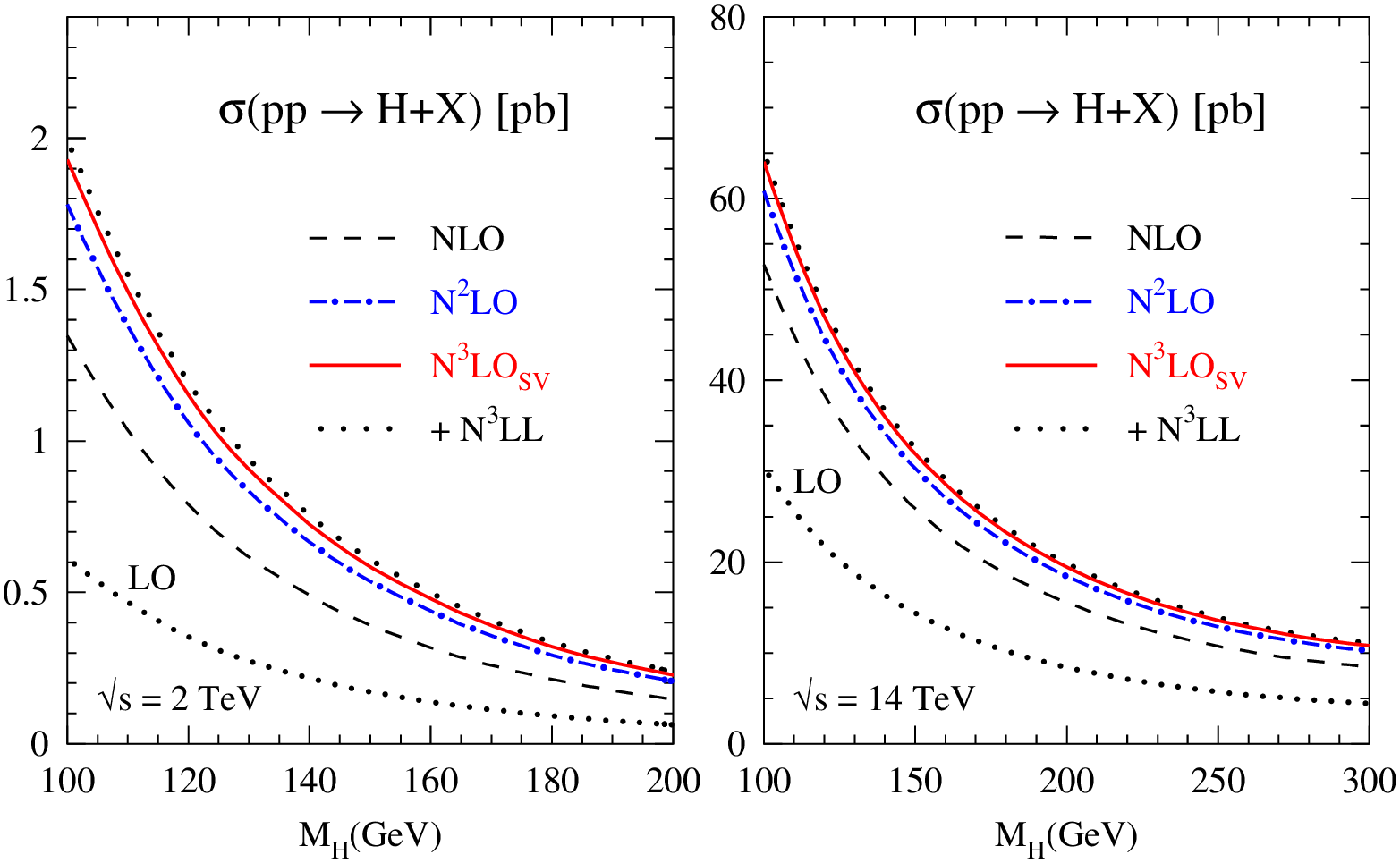}
    \end{tabular}
      \caption[]{\label{fig::n3lo}\sloppy
	Inclusive Higgs production cross section through gluon fusion,
	including the soft \nklo{3} and corresponding resummation
	effects. From Ref.~\cite{Moch:2005ky}.
        }
  \end{center}
\end{figure}

The most recent achievement concerning the fixed-order calculation is
the evaluation of the soft terms through \nklo{3} or, in terms of
\eqn{eq::soft}, the coefficients $b_k$, $k=0,\ldots,5$, through
$\alpha_s^3$~\cite{Moch:2005ky}. The $\delta(1-x)$ piece in
\eqn{eq::soft} receives contributions from the virtual terms which are
still unknown, but one may deduce from the lower order results that they
are numerically small. Note also that the terms for $k=1,\ldots,5$ could
be derived from the {\abbrev NNLL} resummation
formula~\cite{Catani:2003zt}, thus providing a useful check.  The effect
of these \nklo{3} terms is again a mild increase of the cross section
(for $\muF=\muR\approx M_H$), and a reduction of the scale
uncertainty. On the other hand, these terms allow to push the
resummation of the soft terms to higher
orders~\cite{Moch:2005ky,Ravindran:2006cg}, with again rather small
numerical impact (see \fig{fig::n3lo}).

It remains to be said that apart from the phenomenological significance
of such higher order results, they appear to reveal interesting
structures of the perturbative series that are worth studying for their
own sake~\cite{Blumlein:2005im,Dokshitzer:2005bf}.

Clearly, if so much effort is put into minimizing the theoretical
uncertainty due to \qcd{} effects, one needs to start thinking about
electro-weak corrections as well. In fact, terms of order $\GFermi
m_t^2$ have been known for quite a while now, resulting in effects below
1\% of the \lo{} rate~\cite{Djouadi:1994ge}. The full set of Feynman
diagrams can be divided into those that contain a top quark, and those
that contain only light quarks. The latter set has been evaluated in
Ref.\,\cite{Aglietti:2004nj}, while the full result for the former was
obtained in Ref.\,\cite{Degrassi:2004mx}.  The numerical impact of the
electro-weak corrections ranges between 5 and 8\% of the \lo{} term.

{\it Transverse momentum and rapidity distributions.}\quad
A large $K$-factor for the total cross section leaves open the question
on how the radiative corrections affect different regions of phase
space.  Better insight into this issue is provided by differential
quantities such as $\dd\sigma/\dd p_T\dd y$, where $p_T$ and $y$ are the
transverse momentum and the rapidity of the Higgs boson, respectively.

At the partonic level, the Higgs boson can only be produced at finite
transverse momentum $p_T$ if the latter is balanced by the real
radiation of a quark or a gluon. Thus, distributions at non-zero $p_T$
are related to the process $H+$jet whose \lo{} prediction is of order
$\alpha_s^3$. The corresponding \nlo{} effects were studied both
analytically \cite{Glosser:2002gm,Ravindran:2002dc}, and in the form of
a partonic Monte Carlo program~\cite{deFlorian:1999zd}, yielding a
rather flat dependence of the $K$-factor on $p_T$ and $y$ at
intermediate values of these variables.  At small $p_T$, the fixed-order
perturbative approach breaks down, but this can be accounted for by
resummation of logarithms (see, e.g., Ref.\,\cite{Assamagan:2004mu} and
references therein).  At large $p_T$, on the other hand, one again
encounters similar logarithms as for the inclusive rate, arising from
soft gluon radiation. Their resummation is known through {\abbrev
NLL}~\cite{deFlorian:2005rr}.

{\it {\abbrev\it NNLO} Monte Carlo.}\quad
Currently the most general higher order prediction for the gluon fusion
process is available in the form of the partonic \nnlo{} Monte Carlo
program {\tt FEHiP}~\cite{Anastasiou:2005qj}.  It allows to study
arbitrary kinematical distributions of the Higgs boson with \nnlo{}
accuracy, as well as the application of phase space cuts. This provides
detailed information on how the radiative corrections affect the various
regions of phase space.

For example, \tab{tab::fehip} compares the ratio $K^{(2)}$ of the
\nnlo{} to the \nlo{} cross section as obtained by the fully inclusive
calculation (subscript ``inc'') to the one where ``standard cuts'' are
applied (subscript ``cut''; see Ref.\,\cite{Anastasiou:2005qj} for
details). The numbers show that the radiative corrections are only
slightly affected by the cuts, and that the cross section including cuts
is approximated to better than 5\% by the quantity $K_{\rm
inc}^{(2)}\times\sigma^{\rm cut}_{\rm \nlo{}}$.

\begin{table}[htb]
\begin{center}
$$
\begin{array}{||c|c|c||}
\hline M_H,~{\rm GeV} & \sigma^{\rm cut}_{\rm NNLO}/\sigma^{\rm
 inc}_{\rm NNLO} 
 & K^{(2)}_{\rm cut}/K^{(2)}_{\rm inc} \\ \cline{1-3}
110&  0.590 & 0.981 \\ \cline{2-3}
115&  0.597 & 0.968 \\ \cline{2-3}
120&  0.603 & 0.953 \\ \cline{2-3}
125&  0.627 & 0.970 \\ \cline{2-3}
130&  0.656 & 1.00  \\ \cline{2-3}
135&  0.652 & 0.98  \\ \cline{2-3}
\hline
\end{array}
$$
\vspace*{0.5cm}
\caption[]{\label{tab::fehip}
Comparisons between the cut and inclusive cross sections for different 
Higgs masses.  The second column contains the ratio of the NNLO cross section 
with the standard cuts over the inclusive cross section, while the third 
column contains the ratio of cut and inclusive results for the
$K$-factor $K^{(2)} = \sigma_{\rm NNLO} / \sigma_{\rm NLO}$.
It is $\muR=\muF = M_H/2$. From Ref.\,\cite{Anastasiou:2005qj}.}
\end{center}
\end{table}

This observation motivates another step towards more realistic higher
order event simulations: In order to transfer the purely partonic
\nnlo{} result of Ref.\,\cite{Anastasiou:2005qj} to truly hadronic final
states, the {\tt Pythia}~\cite{Sjostrand:2001yu} and {\tt
MC@NLO}~\cite{Frixione:2002ik,Frixione:2003ei} event generators have been
supplemented by a re-weighting grid in the $p_T$-$y$
plane~\cite{Davatz:2006ut}, evaluated from the ratio of the partonic
{\tt FEHiP} and the hadronic results, integrated over two-dimensional
intervals.  This procedure is based on the fact that sufficiently
inclusive quantities should be described equally well in a partonic and
a hadronic approach. A similar strategy was followed in
Ref.\,\cite{Davatz:2004zg}, where the re-weighting was based only on the
\nlo{} $p_T$-spectrum of the Higgs boson, however.

{\it Background calculations.}\quad An important issue for Higgs
searches and studies at the \lhc{} is the theoretical control of
background processes. The number of higher order results available in
this context is way too large to even attempt giving proper credit to
each one of them.  An extensive list of programs to evaluate higher
order cross sections can be found at Ref.\,\cite{hepcode}.

Quite often, side band subtractions rather than \nlo{} simulations will
be the most efficient way in order to separate the signal from the
background, once enough data are available. But in certain cases, such a
procedure will not be possible, for instance if missing energy in the
Higgs decay does not allow the reconstruction of a Higgs mass peak.  An
example for such a case is the $WW$ decay mode of the Higgs boson.  In
fact, in order to use this channel as a discovery mode, one needs to
keep track of the angular correlations among the Higgs decay
products~\cite{Dittmar:1996ss}.  Also here, the \nlo{} corrections have
been known for a while. However, recently it was found that the $gg$
initiated component, although being formally of \nnlo{}, can amount to
30\% of the \nlo{} rate, once the relevant cuts for Higgs searches are
applied~\cite{Binoth:2005ua,Duhrssen:2005bz}.

\section{Weak boson fusion}
The weak boson fusion (\wbf{}) process itself is under very good
theoretical control: the \nlo{} corrections have a comparatively simple
structure, since single gluon exchange between the incoming quarks is
not allowed by color conservation, meaning that 5-point functions are
absent in the calculation. The \nlo{} corrections are available in the
form of a partonic Monte Carlo program~\cite{Figy:2003nv}, allowing for
application of cuts as it is particulary important for this process in
order to separate it from the background.

The challenge concerning radiative corrections is indeed related to
these background processes. The dominant source is Higgs production in
gluon fusion, when the Higgs is associated with two jets. The \lo{}
prediction is of order $\alpha_s^4$, meaning that the renormalization
scale dependence is rather large.  The full top mass dependence of the
cross section at \lo{} has been evaluated in
Ref.\,\cite{DelDuca:2001fn}. The \nlo{} calculation
involves massive two-loop five-point functions and is certainly out of
reach at the moment. However, the \lo{} calculation revealed that for
jet transverse momenta $p_{Tj}\lesssim m_t$, one may integrate out the
top quark, thus arriving at one-loop five-point functions, calculated in
Ref.~\cite{Ellis:2005qe}.  However, these still need to be supplemented
by the real radiation contribution, the amplitude for which has been
evaluated in Ref.\,\cite{DelDuca:2004wt,Dixon:2004za,Badger:2004ty}.

Other important background processes to \wbf{} are $Vjj$ and $VVjj$
production, and also here \nlo{} corrections are
available~\cite{Oleari:2003tc,Jager:2006cp:2006zc}.

\section{Supersymmetry}

Gluon fusion remains one of the most important production modes also in
supersymmetric models. In fact, since the pseudo-scalar Higgs boson in
the \mssm{} has no tree-level coupling to vector bosons, it cannot be
produced through {\abbrev WBF}, for example. Therefore, gluon fusion and
associated $b\bar bA$ production are the dominant production modes in
this case (see, e.g., Ref.\,\cite{Belyaev:2005ct}).

Many of the higher order results that are available for \sm{}
Higgs production can be taken over to the case of the neutral, {\abbrev
CP}-even Higgs bosons within the \mssm{}.  For example, if the squarks
are heavy, they do not contribute significantly to the gluon-Higgs
coupling, which is then again mediated predominantly by top and, for
not too small values of $\tan\beta$, bottom loops. The cross section for
$h,H$-production including \qcd{} corrections can then be derived easily
from the \sm{} expression.

Pseudo-scalar Higgs production, on the other hand, requires a
modification of the top-Higgs and thus the effective gluon-Higgs
coupling with respect to \eqn{eq::leff}. But also here, many higher
order corrections are known, for example the inclusive \nnlo{} cross
section~\cite{Harlander:2002vv,Anastasiou:2002wq,Ravindran:2003um}, and
various \nlo{}
distributions~\cite{Field:2002pb,Field:2004tt,Field:2003yy}.

Larger values of $\tan\beta$ increase the importance of bottom loops to
the gluon-Higgs coupling (see, e.g.,
Refs.\,\cite{Spira:1997dg,Harlander:2003xy}). An effective theory for
the gluon-Higgs interaction along the lines of \eqn{eq::leff} is not
known in this case, such that higher order calculations are much more
difficult. In fact, only the \nlo{} result is available at the moment,
in the form of a one-dimensional integral
representation~\cite{Spira:1995rr}.  Let us remark, however, that the
virtual corrections have meanwhile been expressed in terms of analytic
functions~\cite{Harlander:2005rq}.

If their masses are not too large, top squarks may influence the
gluon-Higgs coupling as well. In this case, the \nlo{} gets more
involved as compared to the \sm{}, mostly because several mass scales
enter the problem. However, one may again employ an effective theory
approach analogous to \eqn{eq::leff}, where top quarks and squarks are
considered as heavy~\cite{Dawson:1996xz,Harlander:2003bb:2004tp}. The
particle spectrum of the effective theory is then the same as in the
\sm{} case, and the only unknown quantity is the Wilson coefficient
$C(\alpha_s)$. The latter is obtained from massive tadpole integrals
which can be evaluated analytically through two loops using the proper
reduction formulas~\cite{Davydychev:1992mt}.

This allows one to evaluate the \nlo{} corrections to \susy{} Higgs
production for small values of $\tan\beta$ (for large $\tan\beta$, the
bottom and sbottom loop effects may be taken into account approximately
using the leading order expression and resummation of $\tan\beta$
terms~\cite{Carena:2000uj}). The results for both the production of a
{\abbrev CP}-even~\cite{Harlander:2003bb:2004tp} and a
{\abbrev CP}-odd~\cite{Harlander:2005if} Higgs boson have been evaluated
through \nlo{} in this way. For the {\abbrev CP}-even Higgs, even an
estimate at \nnlo{} has been obtained~\cite{Harlander:2003kf} (denoted
\nnlo{}' in \fig{fig::sig-c1maxcol} below).

A particulary dramatic scenario is given by the so-called ``gluophobic
Higgs''~\cite{Djouadi:1998az,Carena:1999xa}, where the quark and squark
loops interfere distructively, such that the Higgs coupling to gluons
becomes very small. \fig{fig::sig-c1maxcol} shows the effects of higher
orders in $\alpha_s$ in this region of \susy{} parameter space. One
observes that the radiative corrections do not change the general
behaviour of the cross section.  Rather, the \lo{} rate is multiplied by
an almost constant $K$-factor close to the one of the \sm{}
calculation~\cite{Harlander:2003bb:2004tp}.

\begin{figure}
  \begin{center}
    \begin{tabular}{c}
      \includegraphics[bb=110 255 465
      550,width=.5\textwidth]{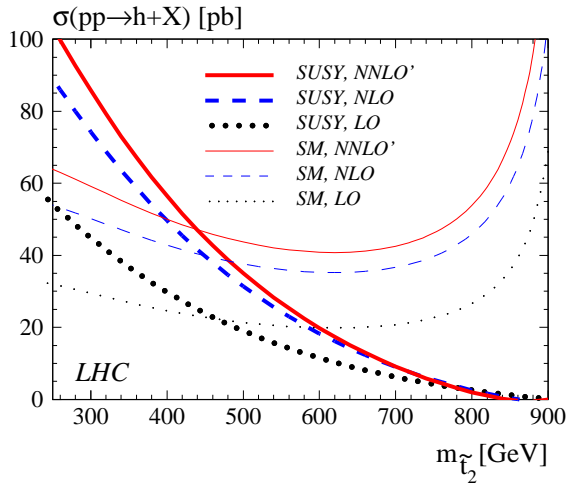}
    \end{tabular}
      \caption[]{\label{fig::sig-c1maxcol}\sloppy Thick lines: gluon
	fusion cross section in the \mssm{} (for details, see
	Ref.\,\cite{Harlander:2003bb:2004tp}). The thin lines show the result
	when the squark effects are neglected.}
  \end{center}
\end{figure}

Concerning differential distributions, the \lo{} diagrams to Higgs plus
jet production due to squark loops have been evaluated in
Refs.~\cite{Brein:2003df,Field:2003yy}. For higher order effects, one
currently needs to rely on the effective Lagrangian approach due to the
complexity of the calculation. Then, however, the \susy{} effects
factorize into the Wilson coefficient, just like for the inclusive rate.

\section{Bottom quark annihilation}

In \susy{}, Higgs production in association with bottom quarks can give
a significant contribution to the total Higgs production cross section
at the \lhc{}. In fact, it can even exceed the gluon fusion component.
The proper theoretical description has been a subject of discussion for
quite some time now. The difficulties arise from the fact that a
potentially large mass difference between the Higgs boson and the bottom
quark can lead to large logarithms that originate from integration over
the collinear region of one or both of the produced bottom quarks.  It
has been suggested to use bottom quark parton densities as a way to
resum these logarithms. However, this so-called 5-flavor scheme
(\ffs{5}) and the 4-flavor scheme (\ffs{4}), where these logarithms are
not resummed, lead to considerably different numerical results for the
total cross section.

It was then realized that the discrepancies between the two approaches
are much smaller once the factorization scale $\muF$ for the bottom
densities is chosen significantly lower than the supposedly ``natural''
choice $\muF=M_H$. In fact, based on the argument that factorization
works only in the collinear limit, it was suggested that a reasonable
choice was $\muF=M_H/4$ in this
case~\cite{Plehn:2002vy,Maltoni:2003pn,Boos:2003yi}. This is because for
$p_T\gtrsim M_H/4$, the $p_T$-distribution of the bottom quarks in the
final state begins to deviate significantly from the collinear form
$\dd\sigma/\dd p_T\sim 1/p_T$~\cite{Rainwater:2002hm}.

\begin{figure}
  \begin{center}
    \begin{tabular}{cc}
      \includegraphics[bb=110 280 465
      560,width=.45\textwidth]{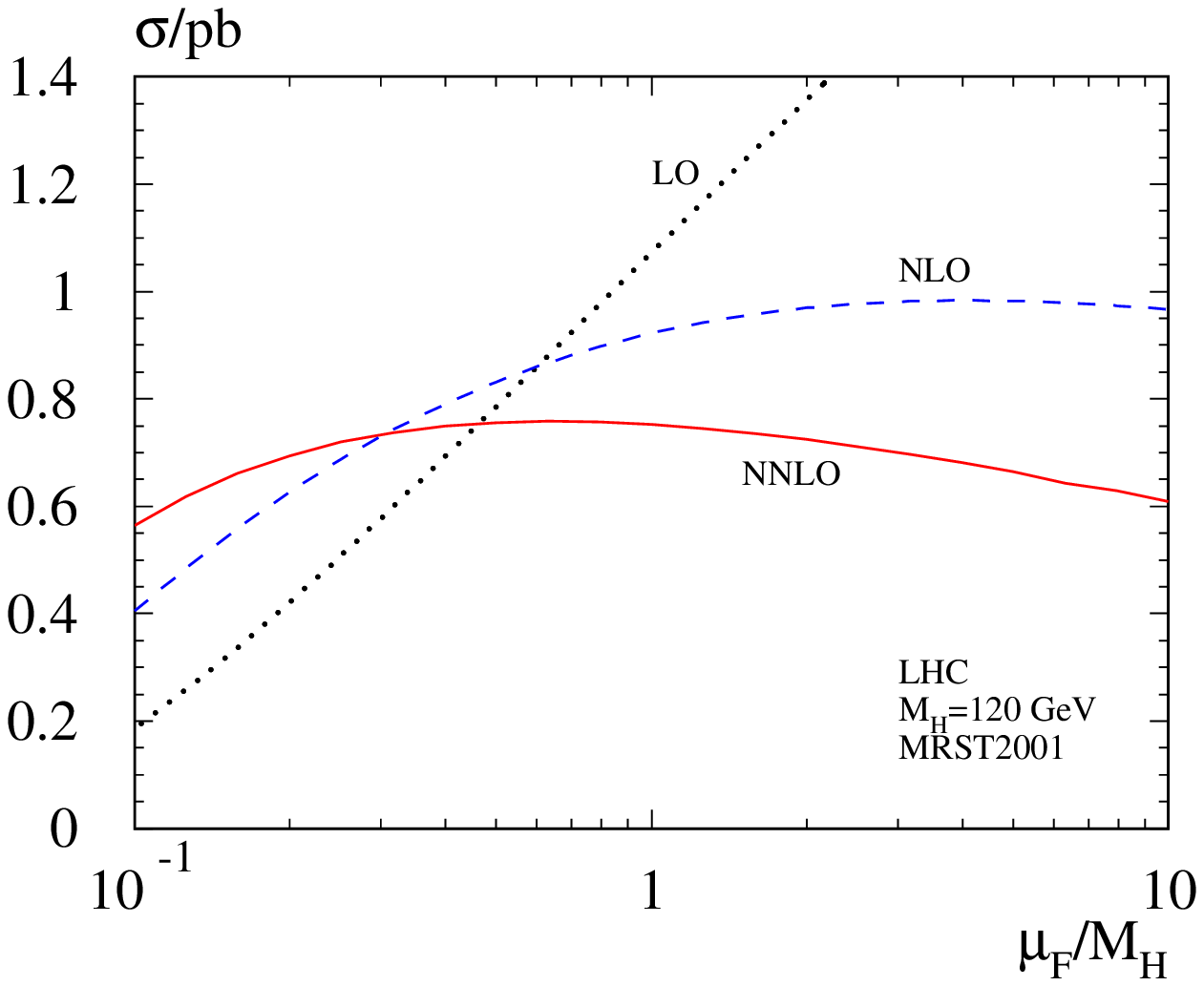} &\quad
      \includegraphics[bb=110 280 465
      560,width=.45\textwidth]{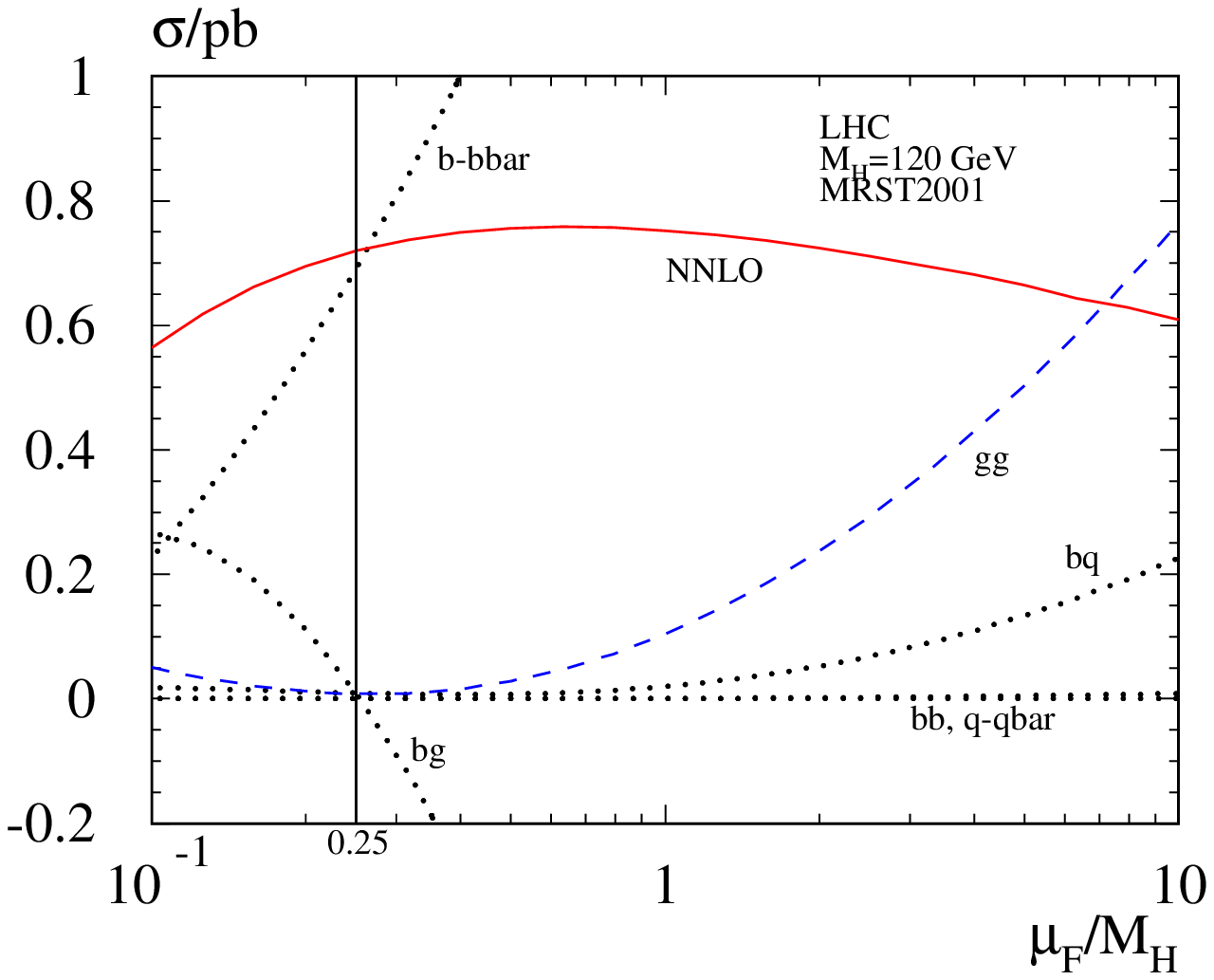} \\
      (a) & (b)
    \end{tabular}
      \caption[]{\label{fig::ggmuf120}\sloppy (a) \lo{}, \nlo{}, and
	\nnlo{} prediction for the inclusive $b\bar bH$ cross section in
	the \ffs{5}~\cite{Harlander:2003ai}.\quad (b) Individual
	components to the \nnlo{} prediction arising from various
	subprocesses in the \msbar{} scheme. See also
	Ref.~\cite{Buttar:2006zd}}.
  \end{center}
\end{figure}

This choice for the factorization scale later has found support from the
\nnlo{} result for that process, evaluated in the
\ffs{5}~\cite{Harlander:2003ai}.  This is shown in
\fig{fig::ggmuf120}~(a) for the \lhc{}. Clearly, convergence of the
perturbative series appears to be much better for scales below $M_H$
rather than above.

Yet another intriguing feature of this \nnlo{} result indicates that
indeed the scale $\muF=M_H/4$ is markedly different from any other. To
see this, \fig{fig::ggmuf120}~(b) shows separately the contributions
from the $b\bar b$, $bg+\bar bg$, $gg$, and other, much smaller partonic
sub-processes in the \msbar{} scheme.  Also shown is the sum of all
these curves (solid line), corresponding to the \nnlo{} result.  It so
happens that right at $\muF=M_H/4$, all contributions except for the
$b\bar b$ term practically vanish simultaneously~\cite{Buttar:2006zd}.

One may wonder whether the bottom quarks can be taken massless for the
$gg$ component of the \nnlo{} contribution, which does not have an
initial state bottom quark. These effects, however, are expected to be
of order $m_b^2/M_H^2$ and thus negligible. This is indeed observed when
comparing the massive~\cite{Maltoni:2003pn} to the
massless~\cite{Harlander:2003ai} result of this
component~\cite{Buttar:2006zd}.

Clearly, the pure \ffs{5} can not be applied directly to {\it exclusive}
$b\bar bH$ production, i.e., if one or both of the bottom jets are
required to be produced at large transverse momenta. In the fully
exclusive case, the \nlo{} corrections are
available~\cite{Dittmaier:2003ej,Dawson:2003kb}. If only one bottom
quark is required at large $p_T$, one may apply the \ffs{5} for the one
that is integrated over. Also in this case, the \nlo{} corrections are
known~\cite{Dicus:1998hs,Campbell:2002zm}.

Concluding this section, the $b\bar bH$ process has been a very
inspiring subject over the past few years, and there would still be
several aspects to be discussed. However, in this short write-up, we
have to refer the interested reader to the recent literature (see, e.g.,
Refs.\,\cite{Buttar:2006zd,Kramer:2004ie,Maltoni:2005wd} and references
therein).

\section{Conclusions}

Higgs physics has been a very fruitful field of research and, with the
\lhc{} data to come, will be even more so in the future.  Progress has been
fast-paced, and many of the results and techniques are general enough to
find applications also in very different contexts.  I have tried to
summarize the most significant developments of the past few years
related to Higgs production at the \lhc{}, and to direct the reader to
the relevant literature whenever more detailed information is required.

{\it Acknowledgments.}\quad I am grateful to the organizers of {\abbrev
WHEPP-9} for the invitation to this inspiring workshop.  In addition, I
would like to thank P.~Agrawal, B.~Allanach, R.~Godbole, S.~Kraml,
P.~Mathews, D.~Miller, M.~M\"uhlleitner, H.~P\"as, V.~Ravindran,
K.~Sridhar and all other participants for the pleasant atmosphere during
the workshop.

\def\app#1#2#3{{\it Act.~Phys.~Pol.~}\jref{\bf B #1}{#2}{#3}}
\def\apa#1#2#3{{\it Act.~Phys.~Austr.~}\jref{\bf#1}{#2}{#3}}
\def\annphys#1#2#3{{\it Ann.~Phys.~}\jref{\bf #1}{#2}{#3}}
\def\cmp#1#2#3{{\it Comm.~Math.~Phys.~}\jref{\bf #1}{#2}{#3}}
\def\cpc#1#2#3{{\it Comp.~Phys.~Commun.~}\jref{\bf #1}{#2}{#3}}
\def\epjc#1#2#3{{\it Eur.\ Phys.\ J.\ }\jref{\bf C #1}{#2}{#3}}
\def\fortp#1#2#3{{\it Fortschr.~Phys.~}\jref{\bf#1}{#2}{#3}}
\def\ijmpc#1#2#3{{\it Int.~J.~Mod.~Phys.~}\jref{\bf C #1}{#2}{#3}}
\def\ijmpa#1#2#3{{\it Int.~J.~Mod.~Phys.~}\jref{\bf A #1}{#2}{#3}}
\def\jcp#1#2#3{{\it J.~Comp.~Phys.~}\jref{\bf #1}{#2}{#3}}
\def\jetp#1#2#3{{\it JETP~Lett.~}\jref{\bf #1}{#2}{#3}}
\def\jhep#1#2#3{{\small\it JHEP~}\jref{\bf #1}{#2}{#3}}
\def\mpl#1#2#3{{\it Mod.~Phys.~Lett.~}\jref{\bf A #1}{#2}{#3}}
\def\nima#1#2#3{{\it Nucl.~Inst.~Meth.~}\jref{\bf A #1}{#2}{#3}}
\def\npb#1#2#3{{\it Nucl.~Phys.~}\jref{\bf B #1}{#2}{#3}}
\def\nca#1#2#3{{\it Nuovo~Cim.~}\jref{\bf #1A}{#2}{#3}}
\def\plb#1#2#3{{\it Phys.~Lett.~}\jref{\bf B #1}{#2}{#3}}
\def\prc#1#2#3{{\it Phys.~Reports }\jref{\bf #1}{#2}{#3}}
\def\prd#1#2#3{{\it Phys.~Rev.~}\jref{\bf D #1}{#2}{#3}}
\def\pR#1#2#3{{\it Phys.~Rev.~}\jref{\bf #1}{#2}{#3}}
\def\prl#1#2#3{{\it Phys.~Rev.~Lett.~}\jref{\bf #1}{#2}{#3}}
\def\pr#1#2#3{{\it Phys.~Reports }\jref{\bf #1}{#2}{#3}}
\def\ptp#1#2#3{{\it Prog.~Theor.~Phys.~}\jref{\bf #1}{#2}{#3}}
\def\ppnp#1#2#3{{\it Prog.~Part.~Nucl.~Phys.~}\jref{\bf #1}{#2}{#3}}
\def\rmp#1#2#3{{\it Rev.~Mod.~Phys.~}\jref{\bf #1}{#2}{#3}}
\def\sovnp#1#2#3{{\it Sov.~J.~Nucl.~Phys.~}\jref{\bf #1}{#2}{#3}}
\def\sovus#1#2#3{{\it Sov.~Phys.~Usp.~}\jref{\bf #1}{#2}{#3}}
\def\tmf#1#2#3{{\it Teor.~Mat.~Fiz.~}\jref{\bf #1}{#2}{#3}}
\def\tmp#1#2#3{{\it Theor.~Math.~Phys.~}\jref{\bf #1}{#2}{#3}}
\def\yadfiz#1#2#3{{\it Yad.~Fiz.~}\jref{\bf #1}{#2}{#3}}
\def\zpc#1#2#3{{\it Z.~Phys.~}\jref{\bf C #1}{#2}{#3}}
\def\ibid#1#2#3{{ibid.~}\jref{\bf #1}{#2}{#3}}

\newcommand{\jref}[3]{{\bf #1}, #3 (#2)}
\newcommand{\bibentry}[4]{#1, #3.}
\newcommand{\arxiv}[1]{{\tt arXiv:#1}}


\begin{thebibliography}{99}
%
%

\bibitem{Beenakker:2001rj:2002nc}
\bibentry{W.~Beenakker, S.~Dittmaier, M.~Kr\"amer, B.~Pl\"umper, 
M.~Spira, P.M.~Zerwas}
{}
{\prl{87}{2001}{201805}; \npb{653}{2003}{151}}
{\arxiv{hep-ph/0107081}; \arxiv{hep-ph/0211352}}

\bibitem{Dawson:2002tg}
\bibentry{S.~Dawson, L.H.~Orr, L.~Reina, D.~Wackeroth}
{Associated top quark Higgs boson production at the {\abbrev LHC}}
{\prd{67}{2003}{071503}}
{\arxiv{hep-ph/0211438}}

\bibitem{Dawson:2003zu}
\bibentry{S.~Dawson, C.~Jackson, L.H.~Orr, L.~Reina, D.~Wackeroth}
{Associated Higgs production with top quarks at the Large Hadron  Collider:
{\abbrev NLO QCD} corrections}
{\prd{68}{2003}{034022}}
{\arxiv{hep-ph/0305087}}

\bibitem{Figy:2003nv}
\bibentry{T.~Figy, C.~Oleari, D.~Zeppenfeld}
{Next-to-leading order jet distributions for Higgs boson production via
weak-boson fusion}
{\prd{68}{2003}{073005}}
{\arxiv{hep-ph/0306109}}

\bibitem{Harlander:2002wh}
\bibentry{R.V.~Harlander and W.B.~Kilgore}
{Next-to-next-to-leading order Higgs production at hadron colliders}
{\prl{88}{2002}{201801}}
{\arxiv{hep-ph/0201206}}

\bibitem{Anastasiou:2002yz}
\bibentry{C.~Anastasiou and K.~Melnikov}
{Higgs boson production at hadron colliders in {\abbrev NNLO QCD}}
{\npb{646}{2002}{220}}
{\arxiv{hep-ph/0207004}}

\bibitem{Ravindran:2003um}
\bibentry{V.~Ravindran, J.~Smith, W.L.~van Neerven}
{{\abbrev NNLO} corrections to the total cross section for Higgs boson
  production  in hadron hadron collisions}
{\npb{665}{2003}{325}}
{\arxiv{hep-ph/0302135}}

\bibitem{Brein:2003wg}
\bibentry{O.~Brein, A.~Djouadi, R.~Harlander}
{{\abbrev NNLO} {\abbrev QCD} corrections to the Higgs-strahlung
processes at hadron colliders} {\plb{579}{2004}{149}}
{\arxiv{hep-ph/0307206}}

\bibitem{Ciccolini:2003jy}
\bibentry{M.L.~Ciccolini, S.~Dittmaier, M.~Kr\"amer}
{Electroweak radiative corrections to associated W H and Z H production  at
hadron colliders}
{\prd{68}{2003}{073003}}
{\arxiv{hep-ph/0306234}}

\bibitem{Djouadi:2005gi:2005gj}
\bibentry{A.~Djouadi}
{}
{\arxiv{hep-ph/0503172}; \arxiv{hep-ph/0503173}}
{}

\bibitem{Buscher:2005re}
\bibentry{V.~B\"uscher and K.~Jakobs}
{Higgs boson searches at hadron colliders}
{\ijmpa{20}{2005}{2523}}
{\arxiv{hep-ph/0504099}}

\bibitem{Dawson:1990zj}
\bibentry{S.~Dawson}
{Radiative corrections to Higgs boson production}
{\npb{359}{1991}{283}}
{}

\bibitem{Djouadi:1991tk}
\bibentry{A.~Djouadi, M.~Spira, P.M.~Zerwas}
{Production of Higgs bosons in proton colliders: {\abbrev QCD} corrections}
{\plb{264}{1991}{440}}
{}

\bibitem{Graudenz:1992pv}
\bibentry{D.~Graudenz, M.~Spira, P.M.~Zerwas}
{{\abbrev QCD} corrections to Higgs boson production at proton-proton
colliders}
{\prl{70}{1993}{1372}}
{}

\bibitem{Kramer:1996iq}
\bibentry{M.~Kr\"amer, E.~Laenen, M.~Spira}
{Soft gluon radiation in Higgs boson production at the {\abbrev LHC}}
{\npb{511}{1998}{523}}
{\arxiv{hep-ph/9611272}}

\bibitem{Spira:1997dg}
\bibentry{M.~Spira}
{{\abbrev QCD} effects in Higgs physics}
{\fortp{46}{1998}{203}}
{\arxiv{hep-ph/9705337}}

\bibitem{Schroder:2005hy}
\bibentry{Y.~Schr\"oder and M.~Steinhauser}
{Four-loop decoupling relations for the strong coupling}
{\jhep{0601}{2006}{051}}
{\arxiv{hep-ph/0512058}}

\bibitem{Chetyrkin:2005ia}
\bibentry{K.G.~Chetyrkin, J.H.~K\"uhn, C.~Sturm}
{QCD decoupling at four loops}
{\npb{744}{2006}{121}}
{\arxiv{hep-ph/0512060}}

\bibitem{Harlander:2000mg}
\bibentry{R.V.~Harlander}
{Virtual corrections to $g g \to H$ to two loops in the heavy top limit}
{\plb{492}{2000}{74}}
{\arxiv{hep-ph/0007289}}

\bibitem{Harlander:2001is}
\bibentry{R.V.~Harlander and W.B.~Kilgore}
{Soft and virtual corrections to $p p \to H + X$ at {\abbrev NNLO}}
{\prd{64}{2001}{01301}}
{\arxiv{hep-ph/0102241}}

\bibitem{Catani:2001ic} 
\bibentry{S.~Catani, D.~de Florian, M.~Grazzini} 
{Higgs production in hadron collisions: Soft and virtual {\abbrev QCD} 
corrections at {\abbrev NNLO}} 
{\jhep{0105}{2001}{025}}
{\arxiv{hep-ph/0102227}}

\bibitem{Catani:2003zt}
\bibentry{S.~Catani, D.~de Florian, M.~Grazzini, P.~Nason}
{Soft-gluon resummation for Higgs boson production at hadron colliders}
{\jhep{0307}{2003}{028}}
{\arxiv{hep-ph/0306211}}

\bibitem{Moch:2005ky}
\bibentry{S.~Moch and A.~Vogt}
{Higher-order soft corrections to lepton pair and Higgs boson production}
{\plb{631}{2005}{48}}
{\arxiv{hep-ph/0508265}}

\bibitem{Ravindran:2006cg}
\bibentry{V.~Ravindran}
{Higher-order threshold effects to inclusive processes in QCD}
{\arxiv{hep-ph/0603041}}
{}

\bibitem{Blumlein:2005im}
\bibentry{J.~Bl\"umlein and V.~Ravindran}
{Mellin moments of the next-to-next-to-leading order coefficient  functions
for the Drell-Yan process and hadronic Higgs-boson  production}
{\npb{716}{2005}{128}}
{\arxiv{hep-ph/0501178}}

\bibitem{Dokshitzer:2005bf}
\bibentry{Y.L.~Dokshitzer, G.~Marchesini, G.P.~Salam}
{Revisiting parton evolution and the large-x limit}
{\plb{634}{2006}{504}}
{\arxiv{hep-ph/0511302}}

\bibitem{Djouadi:1994ge}
\bibentry{A.~Djouadi and P.~Gambino}
{Leading electroweak correction to Higgs boson production at proton
colliders}
{\prl{73}{1994}{2528}}
{\arxiv{hep-ph/9406432}}

\bibitem{Aglietti:2004nj}
\bibentry{U.~Aglietti, R.~Bonciani, G.~Degrassi, A.~Vicini}
{Two-loop light fermion contribution to Higgs production and decays}
{\plb{595}{2004}{432}}
{\arxiv{hep-ph/0404071}}

\bibitem{Degrassi:2004mx}
\bibentry{G.~Degrassi and F.~Maltoni}
{Two-loop electroweak corrections to Higgs production at hadron colliders}
{\plb{600}{2004}{255}}
{\arxiv{hep-ph/0407249}}

\bibitem{Glosser:2002gm}
\bibentry{C.J.~Glosser and C.R.~Schmidt}
{Next-to-leading corrections to the Higgs boson transverse momentum spectrum
in gluon fusion}
{\jhep{0212}{2002}{016}}
{\arxiv{hep-ph/0209248}}

\bibitem{Ravindran:2002dc}
\bibentry{V.~Ravindran, J.~Smith, W.L.~Van Neerven}
{Next-to-leading order QCD corrections to differential distributions of
Higgs boson production in hadron hadron collisions}
{\npb{634}{2002}{247}}
{\arxiv{hep-ph/0201114}}

\bibitem{deFlorian:1999zd}
\bibentry{D.~de Florian, M.~Grazzini, Z.~Kunszt}
{Higgs production with large transverse momentum in hadronic collisions  at
next-to-leading order}
{\prl{82}{1999}{5209}}
{\arxiv{hep-ph/9902483}}

\bibitem{Assamagan:2004mu}
\bibentry{K.A.~Assamagan {\it et al.}  [Higgs Working Group Collaboration]}
{The Higgs working group: Summary report 2003}
{\arxiv{hep-ph/0406152}}
{}

\bibitem{deFlorian:2005rr}
\bibentry{D.~de Florian, A.~Kulesza, W.~Vogelsang}
{Threshold resummation for high-transverse-momentum Higgs production at the
LHC}
{\jhep{0602}{2006}{047}}
{\arxiv{hep-ph/0511205}}

\bibitem{Anastasiou:2005qj}
\bibentry{C.~Anastasiou, K.~Melnikov, F.~Petriello}
{Fully differential Higgs boson production and the di-photon signal  through
next-to-next-to-leading order}
{\npb{724}{2005}{197}}
{\arxiv{hep-ph/0501130}}

\bibitem{Sjostrand:2001yu}
\bibentry{T.~Sj\"ostrand, L.~L\"onnblad, S.~Mrenna}
{PYTHIA 6.2: Physics and manual}
{\arxiv{hep-ph/0108264}}
{}

\bibitem{Frixione:2002ik}
\bibentry{S.~Frixione and B.~R.~Webber}
{Matching NLO QCD computations and parton shower simulations}
{\jhep{0206}{2002}{029}}
{\arxiv{hep-ph/0204244}}

\bibitem{Frixione:2003ei}
\bibentry{S.~Frixione, P.~Nason, B.~R.~Webber}
{Matching NLO QCD and parton showers in heavy flavour production}
{\jhep{0308}{2003}{007}}
{\arxiv{hep-ph/0305252}}

\bibitem{Davatz:2006ut}
\bibentry{G.~Davatz, F.~St\"ockli, C.~Anastasiou, G.~Dissertori, 
M.~Dittmar, K.~Melnikov, F.~Petriello}
{Combining Monte Carlo generators with next-to-next-to-leading order
calculations: Event reweighting for Higgs boson production at the LHC}
{\arxiv{hep-ph/0604077}}
{}

\bibitem{Davatz:2004zg}
\bibentry{G.~Davatz, G.~Dissertori, M.~Dittmar, M.~Grazzini, F.~Pauss}
{Effective $K$-factors for $g g \to H \to W W \to l \nu l \nu$ at the LHC}
{\jhep{0405}{2004}{009}}
{\arxiv{hep-ph/0402218}}

\bibitem{hepcode} {\tt http://www.cedar.ac.uk/hepcode/}

\bibitem{Dittmar:1996ss}
\bibentry{M.~Dittmar and H.K.~Dreiner}
{How to find a Higgs boson with a mass between 155 GeV to 180 GeV at the
LHC}
{\prd{55}{1997}{167}}
{\arxiv{hep-ph/9608317}}

\bibitem{Binoth:2005ua}
\bibentry{T.~Binoth, M.~Ciccolini, N.~Kauer, M.~Kr\"amer}
{Gluon-induced $WW$ background to Higgs boson searches at the LHC}
{\jhep{0503}{2005}{065}}
{\arxiv{hep-ph/0503094}}

\bibitem{Duhrssen:2005bz}
\bibentry{M.~D\"uhrssen, K.~Jakobs, J.J.~van der Bij, P.~Marquard}
{The process $g g \to W W$ as a background to the Higgs signal at the LHC}
{\jhep{0505}{2005}{064}}
{\arxiv{hep-ph/0504006}}

\bibitem{DelDuca:2001fn}
\bibentry{V.~Del Duca, W.~Kilgore, C.~Oleari, C.~Schmidt, D.~Zeppenfeld}
{Gluon-fusion contributions to H + 2 jet production}
{\npb{616}{2001}{367}}
{\arxiv{hep-ph/0108030}}

\bibitem{Ellis:2005qe}
\bibentry{R.K.~Ellis, W.T.~Giele, G.~Zanderighi}
{Virtual QCD corrections to Higgs boson plus four parton processes}
{\prd{72}{2005}{054018}}
{\arxiv{hep-ph/0506196}}

\bibitem{DelDuca:2004wt}
\bibentry{V.~Del Duca, A.~Frizzo, F.~Maltoni}
{Higgs boson production in association with three jets}
{\jhep{0405}{2004}{064}}
{\arxiv{hep-ph/0404013}}

\bibitem{Dixon:2004za}
\bibentry{L.J.~Dixon, E.W.N.~Glover, V.V.~Khoze}
{MHV rules for Higgs plus multi-gluon amplitudes}
{\jhep{0412}{2004}{015}}
{\arxiv{hep-th/0411092}}

\bibitem{Badger:2004ty}
\bibentry{S.D.~Badger, E.W.N.~Glover, V.V.~Khoze}
{MHV rules for Higgs plus multi-parton amplitudes}
{\jhep{0503}{2005}{023}}
{\arxiv{hep-th/0412275}}

\bibitem{Oleari:2003tc}
\bibentry{C.~Oleari and D.~Zeppenfeld}
{Next-to-leading order {\abbrev QCD} corrections to W and Z production via vector-boson
fusion}
{\prd{69}{2004}{093004}}
{\arxiv{hep-ph/0310156}}

\bibitem{Jager:2006cp:2006zc}
\bibentry{B.~J\"ager, C.~Oleari, D.~Zeppenfeld}
{Next-to-leading order QCD corrections to Z boson pair production via
vector-boson fusion}
{\arxiv{hep-ph/0603177}; \arxiv{hep-ph/0604200}}
{}

\bibitem{Belyaev:2005ct}
\bibentry{A.~Belyaev, A.~Blum, R.S.~Chivukula, E.H.~Simmons}
{The meaning of Higgs: tau+ tau- and gamma gamma at the Tevatron and the
LHC}
{\prd{72}{2005}{055022}}
{\arxiv{hep-ph/0506086}}

\bibitem{Harlander:2002vv}
\bibentry{R.V.~Harlander and W.B.~Kilgore}
{Production of a pseudo-scalar Higgs boson at hadron colliders at 
  next-to-next-to leading order}
{\jhep{0210}{2002}{017}}
{\arxiv{hep-ph/0208096}}

\bibitem{Anastasiou:2002wq}
\bibentry{C.~Anastasiou and K.~Melnikov}
{Pseudoscalar Higgs boson production at hadron colliders in {\abbrev
    NNLO} {\abbrev QCD}}
{\prd{67}{2003}{037501}}
{\arxiv{hep-ph/0208115}}

\bibitem{Field:2002pb}
\bibentry{B.~Field, J.~Smith, M.E.~Tejeda-Yeomans, W.L.~van Neerven}
{NLO corrections to differential cross sections for pseudo-scalar Higgs boson
production}
{\plb{551}{2003}{137}}
{\arxiv{hep-ph/0210369}}

\bibitem{Field:2004tt}
\bibentry{B.~Field}
{Next-to-leading log resummation of scalar and pseudoscalar Higgs boson
differential cross-sections at the LHC and Tevatron}
{\prd{70}{2004}{054008}}
{\arxiv{hep-ph/0405219}}

\bibitem{Field:2003yy}
\bibentry{B.~Field, S.~Dawson, J.~Smith}
{Scalar and pseudoscalar Higgs boson plus one jet production at the LHC
  and Tevatron}
{\prd{69}{2004}{074013}}
{\arxiv{hep-ph/0311199}}

\bibitem{Harlander:2003xy}
\bibentry{R.~Harlander}
{Supersymmetric Higgs production at the Large Hadron Collider}
%
{\arxiv{hep-ph/0311005}}
{\arxiv{hep-ph/0311005}}

\bibitem{Spira:1995rr}
\bibentry{M.~Spira, A.~Djouadi, D.~Graudenz, P.M.~Zerwas}
{Higgs boson production at the {\abbrev LHC}}
{\npb{453}{1995}{17}}
{\arxiv{hep-ph/9504378}}

\bibitem{Harlander:2005rq}
\bibentry{R.~Harlander and P.~Kant}
{Higgs production and decay: Analytic results at next-to-leading order
 QCD}
{\jhep{0512}{2005}{015}}
{\arxiv{hep-ph/0509189}}

\bibitem{Dawson:1996xz}
\bibentry{S.~Dawson, A.~Djouadi, M.~Spira}
{{\abbrev QCD} Corrections to {\abbrev SUSY} Higgs Production: 
The Role of Squark Loops}
{\prl{77}{1996}{16}}
{\arxiv{hep-ph/9603423}}

\bibitem{Harlander:2003bb:2004tp}
\bibentry{R.V.~Harlander and M.~Steinhauser}
{}
{\plb{574}{2003}{258-268}; \jhep{0409}{2004}{066}}
{\arxiv{hep-ph/0307346}; \arxiv{hep-ph/0409010}}

\bibitem{Davydychev:1992mt}
\bibentry{A.I.~Davydychev and J.B.~Tausk}
{Two loop selfenergy diagrams with different masses and the momentum
expansion}%
{\npb{397}{1993}{123}}
{}

\bibitem{Carena:2000uj}
\bibentry{M.~Carena, D.~Garcia, U.~Nierste and C.E.M.~Wagner}
{$b \to s \gamma$ and supersymmetry with large $\tan\beta$}
{\plb{499}{2001}{141}}
{\arxiv{hep-ph/0010003}}

\bibitem{Harlander:2005if}
\bibentry{R.V.~Harlander and F.~Hofmann}
{Pseudo-scalar Higgs production at next-to-leading order SUSY-QCD}
{\jhep{0603}{2006}{050}}
{\arxiv{hep-ph/0507041}}

\bibitem{Harlander:2003kf}
\bibentry{R.V.~Harlander and M.~Steinhauser}
{Effects of {\abbrev SUSY-QCD} in hadronic Higgs production at
next-to-next-to-leading order}
{\prd{68}{2003}{111701}}
{\arxiv{hep-ph/0308210}}

\bibitem{Djouadi:1998az}
\bibentry{A.~Djouadi}
{Squark effects on Higgs boson production and decay at the {\abbrev LHC}}
{\plb{435}{1998}{101}}
{\arxiv{hep-ph/9806315}}

\bibitem{Carena:1999xa}
\bibentry{M.~Carena, S.~Heinemeyer, C.E.M.~Wagner, G.~Weiglein}
{Suggestions for improved benchmark scenarios for Higgs-boson searches at
LEP2}
{\arxiv{hep-ph/9912223}}
{\arxiv{hep-ph/9912223}}

\bibitem{Brein:2003df}
\bibentry{O.~Brein and W.~Hollik}
{MSSM Higgs bosons associated with high-$p_T$ jets at hadron colliders}
{\prd{68}{2003}{095006}}
{\arxiv{hep-ph/0305321}}

\bibitem{Plehn:2002vy}
\bibentry{T.~Plehn}
{Charged Higgs boson production in bottom-gluon fusion}
{\prd{67}{2003}{014018}}
{\arxiv{hep-ph/0206121}}

\bibitem{Maltoni:2003pn}
\bibentry{F.~Maltoni, Z.~Sullivan, S.~Willenbrock}
{Higgs-boson production via bottom-quark fusion}
{\prd{67}{2003}{093005}}
{\arxiv{hep-ph/0301033}}

\bibitem{Boos:2003yi}
\bibentry{E.~Boos and T.~Plehn}
{Higgs-boson production induced by bottom quarks}
{\prd{69}{2004}{094005}}
{\arxiv{hep-ph/0304034}}

\bibitem{Rainwater:2002hm}
\bibentry{D.~Rainwater, M.~Spira, D.~Zeppenfeld}
{Higgs boson production at hadron colliders: Signal and background processes}
{\arxiv{hep-ph/0203187}}
{}

\bibitem{Harlander:2003ai} 
\bibentry{R.V.~Harlander and W.B.~Kilgore}
  {Higgs boson production in bottom quark fusion at
     next-to-next-to-leading order}
     {\prd{68}{2003}{013001}}
    {\arxiv{hep-ph/0304035}}

\bibitem{Buttar:2006zd}
\bibentry{C.~Buttar {\it et al.}}
{Les Houches physics at TeV colliders 2005, standard model, QCD, EW, and
Higgs working group: Summary report}
{\arxiv{hep-ph/0604120}}

\bibitem{Dittmaier:2003ej}
\bibentry{S.~Dittmaier, M.~Kr\"amer, M.~Spira}
{Higgs radiation off bottom quarks at the Tevatron and the LHC}
{\prd{70}{2004}{074010}}
{\arxiv{hep-ph/0309204}}

\bibitem{Dawson:2003kb}
\bibentry{S.~Dawson, C.B.~Jackson, L.~Reina, D.~Wackeroth}
{Exclusive Higgs boson production with bottom quarks at hadron colliders}
{\prd{69}{2004}{074027}}
{\arxiv{hep-ph/0311067}}

\bibitem{Dicus:1998hs}
\bibentry{D.~Dicus, T.~Stelzer, Z.~Sullivan, S.~Willenbrock}
{Higgs boson production in association with bottom quarks at
next-to-leading order}
{\prd{59}{1999}{094016}}
{\arxiv{hep-ph/9811492}}

\bibitem{Campbell:2002zm}
\bibentry{J.~Campbell, R.K.~Ellis, F.~Maltoni, S.~Willenbrock}
{Higgs boson production in association with a single bottom quark}
{\prd{67}{2003}{095002}}
{\arxiv{hep-ph/0204093}}

\bibitem{Kramer:2004ie}
\bibentry{M.~Kr\"amer}
{Associated Higgs production with bottom quarks at hadron colliders}
{\arxiv{hep-ph/0407080}}
{\arxiv{hep-ph/0407080}}

\bibitem{Maltoni:2005wd}
\bibentry{F.~Maltoni, T.~McElmurry, S.~Willenbrock}
{Inclusive production of a Higgs or $Z$ boson in association with heavy
quarks}
{\prd{72}{2005}{074024}}
{\arxiv{hep-ph/0505014}}

\end{thebibliography}
\end{document}